\documentclass[]{aa}
\usepackage{graphicx}

\begin{document}

\title{Optical spectroscopy of isolated planetary mass objects in the
 $\sigma$\,Orionis cluster\thanks{
These observations were collected at the VLT of the 
European Southern Observatories} }

   \author{David Barrado y Navascu\'es
          \inst{1,2}
          Mar\'{\i}a Rosa Zapatero Osorio
          \inst{3}
          \and
          Victor J.S. B\'ejar
          \inst{4}
          \and
          Rafael Rebolo
          \inst{4,5}
          \and
          Eduardo L. Mart\'{\i}n
          \inst{6}
          \and
          Reinhard Mundt
          \inst{2}
          \and
          Coryn A.L. Bailer-Jones,
          \inst{2}
          }

\offprints{D. Barrado y Navascu\'es}

   \institute{Departamento F\'{\i}sica Te\'orica, C-XI-506.
              Universidad Aut\'onoma de Madrid, E-28049 Madrid, Spain\\
              \email{barrado@pollux.ft.uam.es}
         \and
             Max-Planck-Institut f\"ur Astronomie,
             K\"onigstuhl 17,      D--69117 Heidelberg. Germany 
         \and
             Division of Geological \& Planetary Sciences,  
             California Institute of Technology, MS 150-21, 
             Pasadena, CA 91125, USA
         \and
             Instituto de Astrof\'\i{}sica de Canarias, E-38205 La Laguna, 
             Tenerife, Spain
         \and
             Consejo Superior de Investigaciones Cient\'{\i}ficas, CSIC, Spain
         \and
             Institute of Astronomy. University of Hawaii at Manoa. 
             2680 Woodlawn Drive, Honolulu, HI 96822, USA.
             }

   \date{Received; accepted }

   \abstract{We have obtained low resolution optical spectra of 15
     isolated planetary mass objects (IPMOs) in the $\sigma$\,Orionis
     cluster, and derived spectral types by comparison with nearby M
     and L dwarfs.  The spectral types are in the range late M -- mid
     L, in agreement with our expectations based on colors and
     magnitudes for bona fide members. Therefore, most of these
     objects have masses below the deuterium burning limit. About 2/3
     show H$\alpha$ in emission at our spectral resolution.  From our
     spectroscopic and photometric data, we infer that three IPMOs in
     this sample may be binaries with components of similar masses.
     These results confirm that the substellar mass function of the
     $\sigma$\,Orionis cluster, in the form dN/dM, keeps rising in the
     planetary domain.  
     \keywords{giant planet formation -- open clusters and
       associations: individual: $\sigma$\,Orionis -- Stars: brown
       dwarfs } }

  \titlerunning{Isolated planetary mass objects in $\sigma$\,Orionis}

  \authorrunning{Barrado y Navascu\'es et al$.$}

   \maketitle

%
%


\section{Introduction}

Very recently, we have discovered 18 very faint, red objects in the
$\sigma$\,Orionis open cluster ($\sim$5 Myr, 352\,pc), using optical
and infrared photometry (Zapatero Osorio et al$.$ 2000).  If they are
indeed members of the association, their masses would be below 18
M$_{\rm jupiter}$ (1047 M$_{\rm jupiter}$ = 1 M$_\odot$).  Other low
mass objects have been found in the young cluster IC~348 (Najita et
al$.$ 2000) and in the Trapezium (Lucas \& Roche 2000; Lucas et al$.$
2001).  They are quite intriguing, since those with masses below 13
M$_{\rm Jupiter}$ would be unable to sustain any nuclear reaction at
any time (in particular deuterium fusion). Therefore, they would have
masses in the planetary domain and some authors have dubbed them
free-floating planets, non-fusors, or isolated planetary mass objects
(IPMOs). These names rely on the mass of the objects, which can be
estimated from observations and comparisons with models, whereas their
origin or formation mechanism cannot be known for sure.  Note that
these mass values are model dependent and should be taken with some
caveats.  Since the first spectroscopic confirmation of the nature of
some IPMOs in $\sigma$\,Orionis and the Trapezium (Zapatero Osorio et
al$.$ 2000; Lucas et al$.$ 2001), they have become very interesting
for two different reasons: (i) they extend the sequence of low mass
objects beyond the deuterium burning threshold (Saumon et al$.$ 1996;
Chabrier et al$.$ 2000); and (ii) their very existence poses a
challenge to our understanding of how they have been created during
the collapse and fragmentation of molecular clouds, since so far no
model has been able to predict the formation of objects in isolation
in this low mass range (e.g. Bodenheimer 1998). Other alternatives
have been recently suggested, such as the formation in multiple
systems as stellar embryos and the ejection from the system before
they accrete enough material to become stars or brown dwarfs  (Boss
2001; Reipurth 2001; Bate 2001).

Here, we present low resolution optical spectroscopy of isolated
planetary mass candidates in the $\sigma$\,Orionis cluster. In our
analysis, we assume a likely cluster age of 5\,Myr, which relies on
several observational facts combined with theory: (i) the
5\,Myr-isochrone provides the best fit to the location of stellar and
substellar $\sigma$\,Orionis members in color-magnitude diagrams
(B\'ejar et al$.$ 2001), (ii) there is no evidence of lithium
depletion in low mass stellar members indicating that the cluster is
younger than 8\,Myr in order to account for total lithium preservation
(Zapatero Osorio et al$.$ 2001), (iii) high mass star models including
mass loss calculations (Meynet et al$.$ 1994) predict that the central
multiple star ($\sigma$\,Orionis itself, O9.5 V, still burning
hydrogen on the main sequence phase) has to be younger that about
7\,Myr.  Therefore, we believe the age of 5\,Myr is very realistic, being
7\, Myr an upper limit for the age of the association.

   \begin{figure}
   \centering
   \includegraphics[width=8cm]{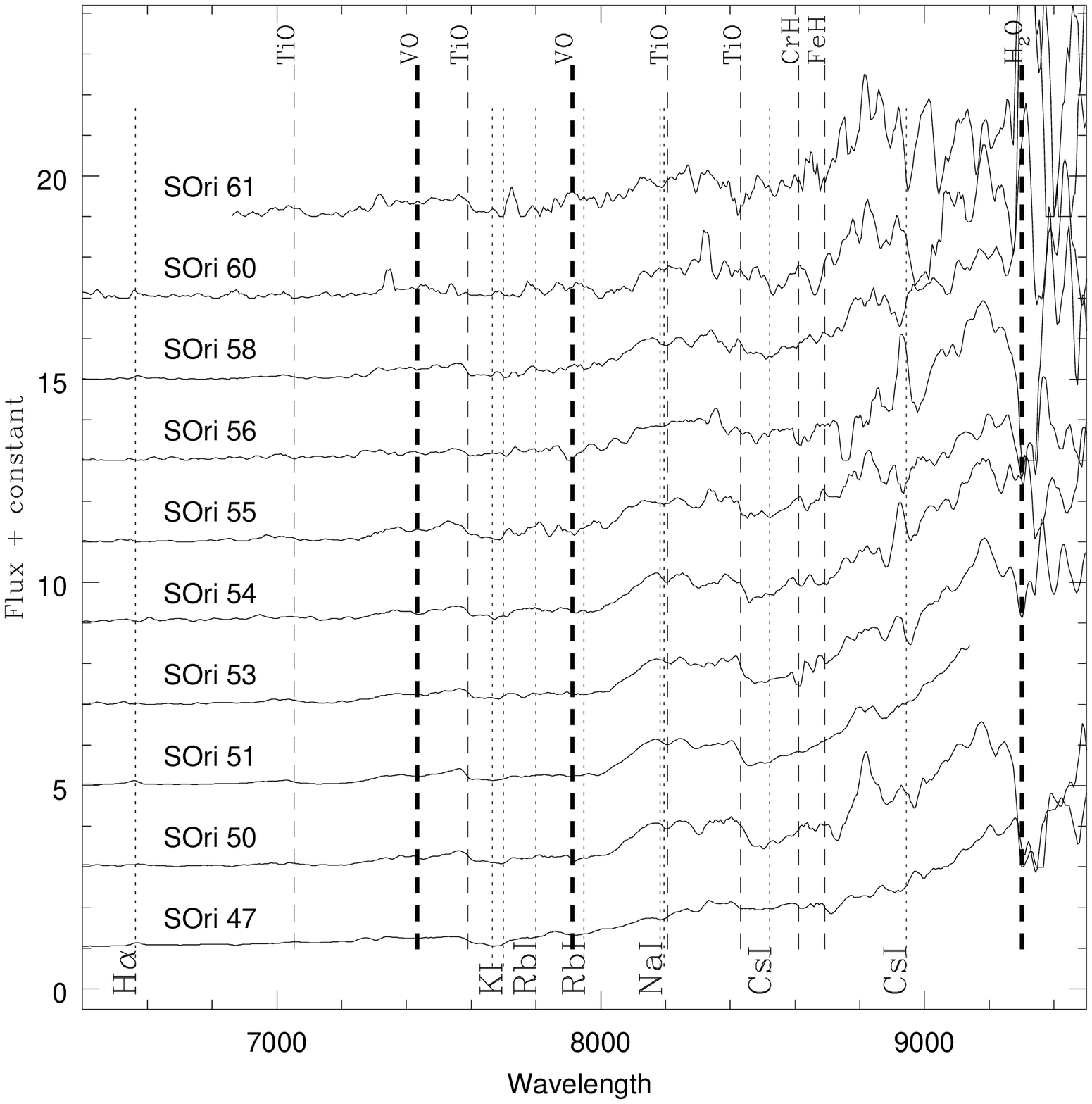}
   \includegraphics[width=8cm]{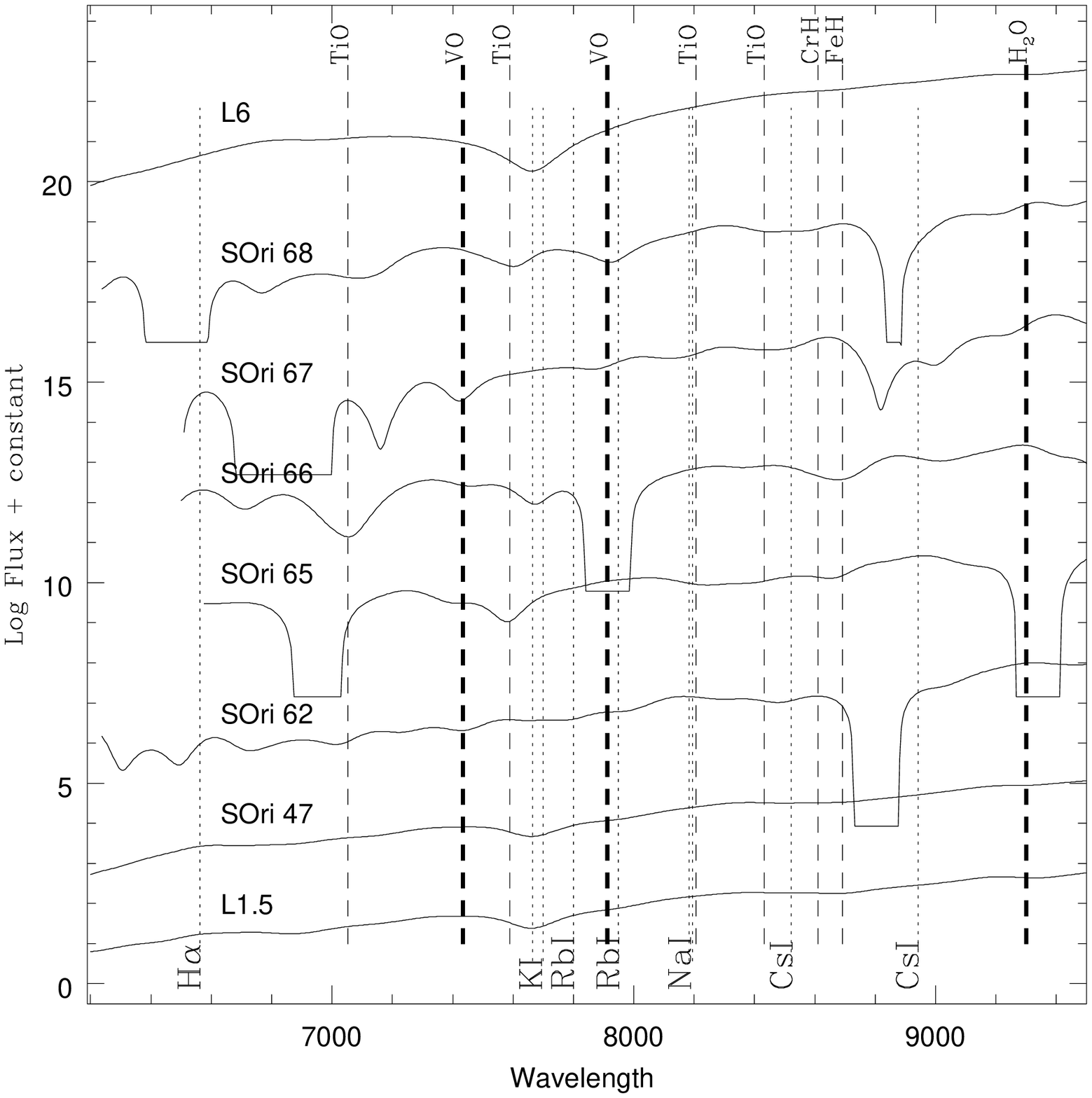}
\caption{VLT/FORS  spectra of isolated 
  planetary mass candidate members of the $\sigma$\,Orionis cluster.
  Top panel: bright objects, data smoothed using a boxcar of 5 pixels.
  Lower panel: faint objects, data smoothed using a boxcar of 5 pixels
  and a Gaussian function with a 12 sigma kernel.  Note the
  logarithmic scale in the y-axis.  For comparison, we include two
  L1.5- and L6-type field objects and S\,Ori\,47.}
         \label{}
   \end{figure}

\begin{table}
\begin{center}
\caption[]{
 $\sigma$\,Orionis  member candidates with derived masses below the 
deuterium burning mass limit (for an age of 5 Myr). }
\begin{tabular}{lcccll}
\hline                                                        
Name       &  $I$    & $I-J$& W$_{\lambda}$(H$\alpha$)& Sp.Type      & Sp.Type         \\
           &         &      & (\AA)     &              & previous        \\
\hline                                                        
SOri47 & 20.53 &  3.15 & 25        & L1.0$\pm$1.0 & L1.5   \\
SOri50 & 20.66 &  3.12 & $<$10     & M9.0$\pm$0.5 &        \\
SOri51 & 20.71 &  3.50 & 25        & M9.0$\pm$0.5 &        \\
SOri53 & 21.17 &  3.28 & $<$10     & M9.0$\pm$0.5 &        \\
SOri54 & 21.29 &  3.30 & 15        & M9.5$\pm$0.5 &        \\
SOri55 & 21.32 &  3.10 &  5        & M9.0$\pm$1.0 &        \\
SOri56 & 21.74 &  3.30 & $<$10     & L1.0$\pm$1.5 & L0.5   \\
SOri58 & 21.90 &  3.30 & 25        & L0.0$\pm$1.0 &        \\
SOri60 & 22.75 &  3.58 & $\sim$25  & L2.0$\pm$0.5 &        \\
SOri61 & 22.78 &  3.16 & --        & L0.0$\pm$1.5 &        \\
SOri62 & 23.03 &  3.59 & $\sim$50  & L2.0$\pm$1.5 & L4.0   \\
SOri65 & 23.23 &  3.33 & $<$20     & L3.5$\pm$2.0 &        \\
SOri66 & 23.23 &  3.40 & $\sim$100 & L3.5$\pm$2.0 &        \\
SOri67 & 23.40 &  3.49 & $\sim$50  & L5.0$\pm$2.0 &        \\
SOri68 & 23.77 &  3.59 & $<$20     & L5.0$\pm$2.0 &        \\
%
%
\hline                                      
\end{tabular}
\end{center}
\end{table}


\section{Observations}

Our spectroscopic data were collected with the Very Large Telescope
Unit \#1 at the Paranal Observatory of the European Southern
Observatory during Dec. 23--27, 2000.  We used the FORS1 spectrograph
and the multi slit capability. FORS1 has a 0.2\arcsec/pixel scale in
the standard resolution, yielding a field of view of
6.8\arcmin$\times$6.8\arcmin.  For our spectroscopic data, we used the
150I grism and the order-blocking filter OG590.  With a slit width of
1.4\arcsec, our resolution is R$\sim$250, as measured in the
comparison arcs.  Our sample of $\sigma$\,Orionis IPMOs was selected
from Zapatero Osorio et al$.$ (2000). In total, we observed 14 out of
the 18 original objects, together with S\,Ori\,47, discovered previously
(Zapatero Osorio et al$.$ 1999).  Table 1 lists magnitudes and colors.
Exposure times are in the range 2400--16800 seconds.

The data were reduced using standard procedures within the IRAF
environment. Individual exposures were added together. Then, the
spectra were extracted using the ``apall'' package within IRAF,
fitting the sky to remove the emission lines and the background.  The
wavelength calibration was performed using HeArHgCd comparison arcs
taken with the same configuration.  Then data were flux calibrated
using spectrophotometric standards.  Finally, we improved the signal
to noise ratio (S/N) by smoothing the spectra with a boxcar of 5
pixels (R$\sim$210). The spectra corresponding to faintest objects
were also convolved with a Gaussian function of 12 sigma kernels
(R$\sim$60).  Figure 1 displays the spectra of our $\sigma$\,Orionis
targets. The upper panel shows the brightest targets (I$_C$=20.5--22.8
mag), with a good signal to noise ratio (S/N=100--40). while the lower
panel displays fainter objects, down to I=23.77 mag, with a much
poorer quality.  Note that when, due to the generally low S/N at the
bottom of the molecular bands, the subtraction of the sky spectrum was
not good enough, we have modified artificially the bad spectral range.
These areas are replaced with horizontal segments in the modified
spectra.  In addition to the $\sigma$\,Orionis targets, we observed
several nearby field objects of M and L spectral type, for comparison
purposes (to derive spectral types).
%


   \begin{figure}
   \centering
   \includegraphics[width=8cm]{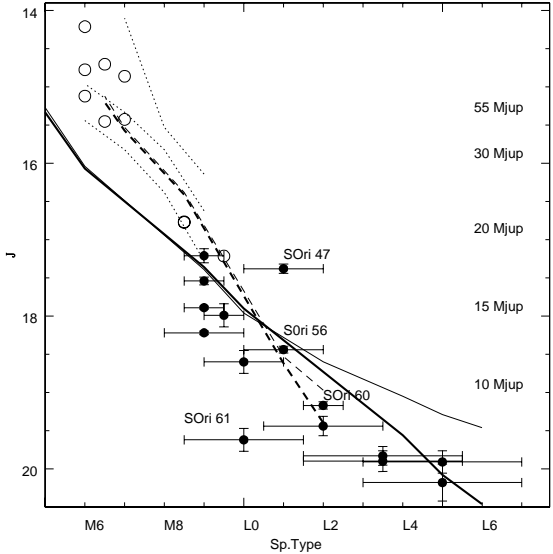}
\caption{ Spectral type against $J$ magnitude.
  Solid circles represent data from this study, whereas open circles
  correspond to data from B\'ejar et al$.$ (2001). The location of the
  brown dwarf--planetary mass domain borderline in $\sigma$\,Orionis
  is at around $J$\,=\,17.9\,mag at the age of 5\,Myr.  The lines
  represent several 5\,Myr  isochrones from Baraffe et al$.$ (1998) --thin lines-
  and Chabrier et al. (2000) --thick lines,  which were
  obtained for  different temperature scales (low gravity by B\'ejar et al. 2001 ---dashed
  lines; high gravity by Basri et al. 2000 ---solid lines, and different gravities 
  by Luhman 1999 ---dotted lines). See text.  }
         \label{}
   \end{figure}


\section{Spectral types}

Figure 1 also depicts several relevant spectral features observable in
the optical, like lines from alkali elements (K\,{\sc i} $\lambda$7665
and $\lambda$7699\,\AA, Na{\,{\sc i} $\lambda$8183 and
  $\lambda$8195\,\AA, Cs\,{\sc i} $\lambda$8521 and
  $\lambda$9843\,\AA, and Rb\,{\sc i} $\lambda$7800 and
  $\lambda$7948\,\AA, dotted vertical lines), and molecular
  absorptions of TiO, FeH, CrH (vertical thin dashed lines), and of VO
  and H$_2$O (vertical thick dashed lines).  In particular, the
  differences can be appreciated, both in the slope of the
  pseudo-continuum as well as the change in the strength and width of
  the K\,{\sc i} resonance doublet, VO and TiO bands, etc.  Detailed
  studies of characteristics of late M and field L very low mass stars
  and brown dwarfs can be found in Kirkpatrick et al$.$ (1999) and
  Mart\'{\i}n et al$.$ (1999).  Figure 1 indicates that, with this
  resolution, it is possible to distinguish key spectral features and
  attempt a spectral classification of the $\sigma$\,Orionis candidate
  members.
  
  Spectral types have been assigned following the scheme proposed in
  Mart\'{\i}n et al$.$ (1999). We have measured flux ratios between
  several bands (pseudo-continuum, VO, TiO, etc) and compared them
  with M- and L-type, nearby field objects. For the faintest object in
  the sample, having a low S/N, we derived a spectral type based on
  the slope of the continuum. Our final spectral types and
  uncertainties are listed in Table~1. Note that the indices we have
  used, in particular PC3 (Mart\'\i n et al$.$ 1996), are based on
  field, older objects, which have higher surface gravity than members
  of $\sigma$ Orionis. Evolutionary models predict gravities about
  log\,$g$\,=\,3.5--4.0 for very low mass objects at young ages around
  5\,Myr. L-type sources (very cool temperatures) display optical
  energy distributions characterized by strong atomic lines of
  Na\,{\sc i} and K\,{\sc i} (Allard et al. 2001; Pavlenko et al.
  2000), which markedly dominate the shape of the far-red wavelengths.
  These lines are very sensitive to gravity (Allard et al. 2001): low
  gravity L-type spectra display significantly less intense alkaline
  lines. Because the blue side of the PC3 index is located very close
  to the K\,{\sc i} resonance doublet, it provides earlier L spectral
  types for low gravity objects than for high gravity ones. The
  M-classes are not so much affected (B\'ejar 2001) by the gravity
  effect. On the contrary, the near-infrared spectra of these types
  are governed by water vapor absorptions, which are steeper for low
  gravities (Allard et al$.$ 2001). Our latest optical spectral types
  appear shifted by 1--2 subclasses towards warmer L-types compared to
  the near-infrared classification given in Mart\'\i n et al$.$ (2001)
  for the objects in common.
  
  The objects in our target list define a neat spectral sequence: the
  fainter the magnitude and redder the color, the cooler the spectral
  class.  We show the relation between spectral type and $J$ magnitude
  in Fig.~2.  Open circles correspond to $\sigma$\,Orionis brown
  dwarfs from B\'ejar et al$.$ (2001), whereas solid circles stand for
  the data studied here.  Estimated masses for each magnitude appear
  on the right-hand side of the diagram on the basis of the 5\,Myr
  dusty isochrone by Chabrier et al$.$ (2000). Albeit this is the
  likely age of the cluster (see Sect.~1), we note that the location
  of the borderline between brown dwarfs and IPMOs 
  ($J \sim$\,17.9\,mag at 5\,Myr) does not
  change by a large amount for other possible ages such as 3\,Myr ($J
  \sim$\,17.5\,mag), 7\,Myr ($J \sim$\,18.2\,mag), and the oldest
  value of 10\,Myr ($J \sim$\,18.6\,mag).
  
  For comparison purposes, Fig.~2 also includes several 5\,Myr isochrones
  using the grainless models of Baraffe et al$.$ (1998)
  and dusty models of Chabrier et al. (2000). These models
  provide magnitudes in the filters of interest and are represented as 
  thin and thick lines, respectively.
  Essentially, there is no difference in this diagram between dusty
  and grainless
  models for objects warmer than about L1.
  We have adopted several 
  calibrations  in order to derive spectral types from effective
  temperatures:\\ 
  (i) Leggett et al$.$ (2000) for early- to mid-M
  types and Basri et al$.$ (2000) for later classes, which appear as 
  solid lines.\\
  (ii)   B\'ejar (2001), shown as dashed lines (see below).\\
  (iii) Luhman (1999), represented as dotted lines. In this last case,
  calibrations for giant, intermediate and dwarf objects have been included
  (from top to bottom).\\
  This last set of isochones shows the effect of gravity: the lower the
  gravity, the more luminous the object is in the J band.
  Leggett et al$.$ (2000) and Basri et al$.$ (2001)
  obtained their temperature scale by studying objects in the field
  and assuming gravities in the range log\,$g$\,=\,5.0--5.5, whereas
  B\'ejar's scale was derived for the lower gravity of
  log\,$g$\,=\,3.5, which is more appropiate for members of the young
  $\sigma$\,Orionis cluster. This late author compared observed low
  resolution optical spectra of cluster M-type brown dwarfs and
  S\,Ori\,47 to spectral synthesis computed for the dusty model
  atmospheres of Allard et al$.$ (2001) following the prescriptions
  given in Pavlenko et al$.$ (2000).  The agreement between
  observations and isochrones of Fig.~2 is better for the B\'ejar's
  (2001) temperature scale. 
  We note that the temperature calibration
  of Luhman (1999), which is determined for gravities intermediate
  between those of dwarfs and giants, also provides a reasonable fit
  to our observations --objects with spectral types earlier than M9--
  and is indeed similar to that of B\'ejar (2001) in this spectral range.
  Finally, dusty models seem to fit better the observations for objects
  having spectral types later than L1.
  In any case, the current uncertainties involved in the conversion
  between effective temperatures and spectral types are still rather
  high and we believe more data are needed to better constrain
  temperature scales for objects of different gravities.
  
  As seen from Fig.~2, there is a clear, monotonic spectral sequence
  from mid-M ($\sigma$\,Orionis massive brown dwarfs) to mid-L
  ($\sigma$\,Orionis IPMOs).  However, there are few exceptions.
  S\,Ori\,61 deviates from the cluster sequence since it shows a
  magnitude fainter than expected for its spectral class.  Thus, we
  cannot confirm its membership in the cluster. On the other hand,
  S\,Ori\,47 clearly stands out among other $\sigma$\,Orionis members.
  The quality of our data, and previously published data by Zapatero
  Osorio et al$.$ (1999, 2000), confirms that this object, with a
  magnitude around the deuterium burning mass limit at the age of the
  cluster, has a spectral type of L1.0--1.5.  Actually, S\,Ori\,47
  could be a nearly equal mass binary comprised of two IPMOs. If this
  is later confirmed with follow-up observations, this object would
  become the first binary IPMO with components of 9--13\,$M_{\rm
    Jupiter}$ each for the age range of 1--7\,Myr.  S\,Ori\,56 and
  S\,Ori\,60 also appear too bright for their spectral types in
  Fig.~2.  Moreover, S\,Ori\,60 shows overluminous and with very red
  colors in infrared color-magnitude diagrams.  Our estimated binarity
  fraction (3 out of 18) in the very low-mass regime of the
  $\sigma$\,Orionis cluster agrees with recent observations of
  binaries among field L-dwarfs (about 20\%, see Reid et al$.$ 2001).
  A search for visual companions around $\sigma$\,Orionis IPMOs using
  $K$-band imaging can be found in Mart\'{\i}n et al$.$ (2001).

   \begin{figure}
   \centering
   \includegraphics[width=8cm]{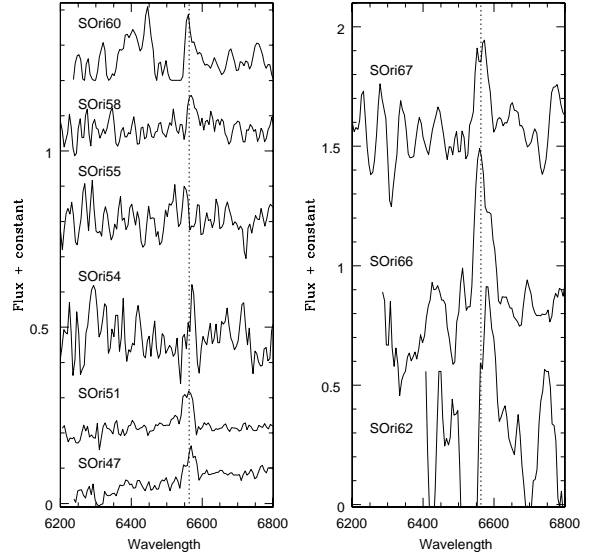}
\caption{Detail around the H$\alpha$ line. No 
smoothing or convolution  with  Gaussian functions
 have been performed.}
         \label{}
   \end{figure}


\section{H$\alpha$, membership and the initial mass function}

The equivalent width of the H$\alpha$ line at $\lambda$6563\AA{ } is
considered as an age indicator in M-dwarfs, and it is usually
associated to stellar activity. In general, the stronger the emission
line for a given spectral type, the younger the object. We have
identified this feature in emission for the first time in such low
mass objects, and measured its equivalent width (EW). The results are
listed in Table~1. For those objects we do not see H$\alpha$ in
emission we can impose an upper limit to the EW at 10\,\AA~ and
20\,\AA, depending on their brightness. Nine objects out of 14 (the
spectrum of one of them does not reach this wavelength) have a
significant emission in H$\alpha$, a strong evidence that they are
young and probable members of the stellar association.  A close-up of
the area around H$\alpha$ is depicted in Fig. 3. We note that very few
H$\alpha$ emissions have been detected in similar spectral type field
objects (Kirkpatrick et al$.$ 1999; Gizis et al$.$ 2000), and the EWs
of the lines are typically below 10\,\AA.  The origin of this feature
is not clear for IPMOs.  Actually, the emission could be due to the
presence of mass accretion from a gas-dust disk (Muzerolle et al$.$
2000).  If this is true this fact might be indicating that these
objects have formed in isolation by direct collapse and cloud
fragmentation, and they are not runaways from embryonic multiple
systems.  A diagram illustrating the H$\alpha$ behaviour as a function
of $(I-J)$ color is presented in Fig. 4. It seems that H$\alpha$
emission is larger for cooler objects.  One of our targets, SOri\,55,
may have experienced a flare-like episode displaying a considerable
variability in the H$\alpha$ EW (Zapatero Osorio et al$.$ 2001).
Moreover, S\,Ori\,47 also has a variable H$\alpha$ emission, since we
measure a EW of 25\,\AA~on the VLT spectrum while Zapatero Osorio et
al$.$ (1999) found an upper limit of 6\,\AA.

   \begin{figure}
   \centering
   \includegraphics[width=8cm]{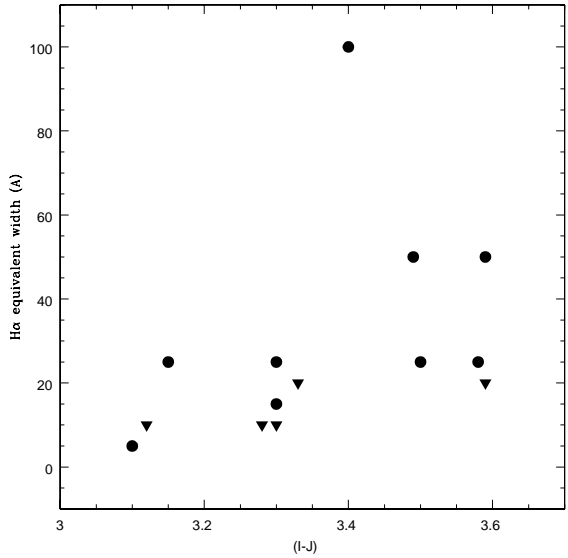}
\caption{H$\alpha$ equivalent width  against the color $(I-J)$. Detections
  and upper limits appear as circles and triangles, respectively. }
         \label{}
   \end{figure}
   
   All the available information, both spectroscopic and photometric,
   indicates that most of the objects in our sample are bona-fide
   members, with masses in the planetary domain and slightly above it.
   Recently, B\'ejar et al$.$ (2001) have derived a initial mass
   function (IMF) for the substellar domain of the $\sigma$\,Orionis
   cluster, finding $\alpha$=0.8$\pm$0.4 (where $dN/dM=kM^{-\alpha}$).
   Our data confirm that the smallest mass bin of that IMF,
   corresponding to planetary mass members, is essentially unaffected
   by photometric contamination, i.e., interlopers. Therefore, the
   cluster IMF keeps rising below the deuterium burning limit,
   implying that the cluster contains a large number of brown dwarfs
   and IPMOs.


\section{Conclusions.}

Using VLT/FORS1 low resolution spectrograph, we have derived spectral
types for 15 isolated planetary mass object (IPMO) candidates in the
$\sigma$\,Orionis cluster discovered by Zapatero Osorio et al$.$
(2000).  All of these objects but one (S\,Ori\,61) appear to be bona
fide members of the cluster, since they follow a well defined sequence
in the magnitude versus spectral type and color diagrams.  A
significant fraction of the IPMOs presents H$\alpha$ in emission, and
at least two of them show variability in this activity indicator
(S\,Ori\,47 and S\,Ori\,55).  Since their membership in the cluster is
confirmed with our optical spectroscopic data, the masses of these
planetary objects must be in the range 18--8\,$M_{\rm Jupiter}$
adopting a cluster age of 5\,Myr.  The combined information provided
by the color-magnitude and the spectral type-magnitude diagrams
indicates that S\,Ori\,47 is a likely photometric binary.  Other two
objects (S\,Ori\,56 and S\,Ori\,60) might also be photometric
binaries.  The spectra presented here prove that membership is
essentially correct in Zapatero Osorio et al$.$ (2000).  Therefore,
the substellar mass function derived by B\'ejar et al$.$ (2001), is
not biased by spurious members in the planetary mass domain.


\begin{acknowledgements}
We thanks the  ESO staff at Paranal Observatory.
The referee, G. Chabrier, has helped to improve the
original version with his  comments and suggestions.
 Partial  financial support was provided by the Spanish
 DGES project PB98--0531--C02--02 
and CICYT grant ESP98--1339-CO2.

\end{acknowledgements}



\begin{thebibliography}{}


\bibitem[2001]{allaerd2001} 
Allard F., Hauschildt P. H., Alexander D. R., Tamanai A., Schweitzer A., 
2001, ApJ, in press


\bibitem[1998]{baraffe98} 
Baraffe I., Chabrier G.,
Allard F., Hauschildt P. H.,
 1998, A\&A, 337, 403

\bibitem[2000]{basri2000} 
Basri G., Mohanty S., Allard F., Hauschildt P. H.,
Delfosse X.  Mart\'{\i}n E. L., Forveille T., Goldman B.,
2000, ApJ 538, 363 


\bibitem[2001]{bate01}
Bate M., 2001, in ``The origins of Stars and Planets: the VLT view'',
eds. J. Alves \& M. McCaughrean,
 Springer-Verlag series "ESO Astrophysics Symposia", in press


\bibitem[2001]{Bejar2001b}  
B\'ejar V. J. S.,  
2001, PhD dissertation, Universidad de La Laguna, Spain

\bibitem[2001]{Bejar2001a}  
B\'ejar V. J. S., Mart\'{\i}n E. L., Zapatero Osorio  M. R., Rebolo, R. 
    Barrado y Navascu\'s D., Bailer-Jones C. A. L., Mundt R., Baraffe 	I,
    Chabrier G., and F. Allard  F., 
2001, ApJ 556, 830

\bibitem[1998]{bod98}
Bodenheimer P. 1998, in ASP Conf. Ser., vol. 134, Brown
Dwarfs and Extrasolar Planets, ed. R. Rebolo, E. L. Mart\'{\i}n, M. R. Zapatero 
Osorio, (San Francisco:ASP), 115

\bibitem[2001]{boss01} 
Boss  A.P.,  2001, ApJ, 551, L167


\bibitem[2000]{Chabrier2000}
Chabrier G.,  Baraffe I., , Allard F., , Hauschildt P., 2000, ApJ  542, L119.

\bibitem[2000]{gizis00} 
Gizis J.E.,  Monet D. G.,Reid I. N., Kirkpatrick J. D.,
Liebert J., Williams R. J.,
 2000, AJ, 120, 1085

\bibitem[1999]{Kirkpatrick1999}
Kirkpatrick D., Reid I. N., Liebert J.,Cutri R. M., Nelson B.,
 Beichman C. A., Dahn C. C., Monet D. G., Gizis J. E., Skrutskie M. F.,
1999, ApJ 519, 802


\bibitem[2000]{leggett00}
Leggett S.K., Allard F. Dahn C., Hauschildt P. H., Kerr T. H., Rayner J.
2000, ApJ 535, 965

\bibitem[2000]{lucas00}
  Lucas  P.\,W., \& Roche, P.\,F. 2000, MNRAS, 314, 858
 
\bibitem[2001]{lucas01}
  Lucas  P.\,W., Roche, P.\,F., Allard, F., \& Hauschildt. 
2000, MNRAS 326, 695

\bibitem[1999]{luhman99}
  Luhman  K.\,L. 1999, ApJ, 525, 466

\bibitem[1996]{martin96}
  Mart\'{\i}n E.\,L.,  Rebolo, R., \& Zapatero Osorio, M.\,R. 1996, ApJ, 469, 706

\bibitem[1999]{martin99}
  Mart\'{\i}n E.\,L., Delfosse X.,  Basri G.,  Goldman B.,  Forveille T.,
 Zapatero Osorio M.R., 
 1999, AJ 118, 2466

\bibitem[2001]{martin2001}
  Mart\'{\i}n E.\,L., Zapatero Osorio M.R., Barrado y Navascu\'es D., 
B\'ejar V., J. S., Rebolo R. 
2001, ApJ Letters,  accepted

\bibitem[2001]{meynet1994}
Meynet,G.,  Maeder A., Schaller G., Schaerer D., Charbonnel C.,
1994, A\&A Suppl. 103, 97


\bibitem[2000]{muzerolle2000}
Muzerolle J., Brice\~no C.,Calvet N., Hartmann L., Hillenbrand L., Gullbring E.,
2001, ApJ Letters 545 L141

\bibitem[2000]{Najita2000}
Najita J., Tiede, G.P., Carr., J.S., 
2000, 541, 977

\bibitem[2000]{Pavlenko2000}
Pavlenko Ya., Zapatero Osorio M. R., Rebolo R.,
2000, A\&A 355, 245

\bibitem[2000]{Reid2001}
Reid I., Gizis J., E., Kirkpatrick J. D.,  Koerner D. W.,
2001, AJ 121, 489

\bibitem[2001]{reipurth01}
 Reipurth B., 2001, in ``The origins of Stars and Planets: the VLT view'',
eds. J. Alves \& M. McCaughrean,
 Springer-Verlag series "ESO Astrophysics Symposia", in press

\bibitem[1996]{Saumon1996} 
Saumon D.,  Hubbard W. B.,Burrows A., Guillot T.,
 Lunine J. I., Chabrier G.,
1996, ApJ 460, 993

\bibitem[1999]{osorio99}
  Zapatero Osorio M.\,R., 
B\'ejar V. J. S., Rebolo R.,
 Mart\'{\i}n, E. L., Basri G.,
 1999, ApJ, 524, L115

  \bibitem[2000]{Zapatero2000} 
 Zapatero Osorio M.R., 
B\'ejar V. J. S.; Mart\'{\i}n E. L.,
Rebolo R.,
 Barrado y Navascu\'es D.,
 Bailer-Jones C. A. L., Mundt R.,
2000, Science, 290, 103

  \bibitem[2001]{Zapatero2001} 
 Zapatero Osorio M.R., et al$.$
2001, in prep.


\end{thebibliography}
\end{document}